\documentclass[%
 aip,
 jmp,%
 amsmath,amssymb,
 preprint,%
prl]{revtex4-1}

\usepackage{graphicx}
\usepackage{dcolumn}
\usepackage{bm}
\usepackage{soul}
\begin{document}

\title[Statistical analysis of HW turbulence]
      {Statistical analysis of Hasegawa - Wakatani turbulence} 

\author{Johan Anderson}%
\email{anderson.johan@gmail.com.}
\affiliation{ 
Department of Earth and Space Sciences, Chalmers University of Technology, 
SE-412 96 G\"{o}teborg, Sweden
}%

\author{Bogdan Hnat}%
\affiliation{ 
Department of Physics, University of Warwick, Coventry, CV4 7AL, United Kingdom
}%

\date{\today}

\begin{abstract}
Resistive drift wave turbulence is a multipurpose paradigm that can be used to understand transport at the edge of fusion devices. The Hasegawa-Wakatani model captures the essential physics of drift turbulence while retaining the simplicity needed to gain a qualitative understanding of this process. We provide a theoretical interpretation of numerically generated probability density functions (PDFs) of intermittent events in Hasegawa-Wakatani turbulence with enforced equipartition of energy in large scale zonal flows and small scale drift turbulence. We find that for a wide range of adiabatic index values the stochastic component representing the small scale turbulent eddies of the flow, obtained from the ARIMA model, exhibits super-diffusive statistics, consistent with intermittent transport. The PDFs of large events (above one standard deviation) are well approximated by the Laplace distribution, while small events often exhibit a Gaussian character. Furthermore there exist a strong influence of zonal flows for example, via shearing and then viscous dissipation maintaining a sub-diffusive character of the fluxes.  
\end{abstract}

\pacs{52.35.Ra, 52.25.Fi, 52.35.Mw, 52.25.Xz}
\keywords{Hasegawa-Wakatani model, Resistive drift waves, stochastic theory, time series analysis}
\maketitle


\section{Introduction}
\label{sec:intro}
The characterization of anomalous transport, that is, transport elevated beyond classical estimates, is an outstanding topic
in fusion plasma research. It is particularly important for the edge region in magnetically confined plasma devices
where turbulence severely limits plasma confinement and is a significant barrier in obtaining fusion \cite{horton1999}.
Strong radial gradients in edge plasmas lead to an elevated anisotropy between parallel and perpendicular length scales
and make the non-adiabatic electron response a crucial component of the system's dynamics.
One paradigm developed to understand plasma turbulence invokes resistive drift waves \cite{horton1999} that access free energy
from the pressure gradient and drive turbulent fluctuations in density ($n$) and electrostatic potential ($\phi$). Physical
understanding of this process has been developed from simplified qualitative models \cite{hm1977, hm1978, hmk1979, hw1983, hw1986}
that include essential features and scalings. The ability to associate certain transport characteristics with particular
physics of the model elucidates experimental results, as well as predictions from quantitative but more complex counterparts
such as gyrokinetic models. One particular example of a simplified model is the Hasegawa-Wakatani (HW) model 
\cite{hmk1979, hw1983, hw1986, dewhurst2008, dewhurst2009, dueck2013} that is in an intermediate regime between adiabatic and
hydrodynamic electrons. This allows for the electrons to dynamically and self-consistently determine the relation between the
density and the electrostatic potential through the turbulence. 

Recently, the need to investigate large scale transport events such as bursts, streamers and blobs have been recognized 
\cite{hi1996, zweben2007, politzer2000, beyer2000, drake1988, antar2001, carreras1996, nagashima2011, dif2010}. These intermittent
events are characterized by a bursty temporal structure while being radially coherent. The Probability Distribution
Functions (PDFs) of fluxes associated with these events have elevated tails compared to a Gaussian distribution, which is a
manifestation of large events or coherent structures mediating transport \cite{politzer2000}. This statistical intermittency
is quantified by higher order cumulants of the PDFs (e.g. skewness and kurtosis). However, a key dynamical feature of
magnetically confined plasma includes a different kind of structure, which is radially localized while extended
in the poloidal direction. These are known as zonal flows. Zonal flows are generated by the small scale turbulence and may act in
a self-regulating manner and govern the saturation of the drift waves \cite{DiamondEA2005, DiamondEA2011, ConnorEA2004, ItohEA2006, 
ConnorMartin2007}.  

A distinguishing feature of turbulent transport is its apparent randomness, complexity at multiple length scales and sensitivity
to initial conditions. Plasma turbulence can be thought of as a nonlinear process with a stochastic component.
In this work, we examine quasi-stationary time series of the electrostatic potential and corresponding vorticity (poloidally
averaged and sampled at different radial points) from the HW simulations. In order to characterize intermittent and random features of the HW turbulence we apply a standard Box-Jenkins modelling for each time series. This mathematical procedure effectively removes deterministic autocorrelations from the time series and renders possible a statistical interpretation of the residual stochastic part. Earlier work using the ARIMA modelling for drift turbulence found that an ARIMA(s,0,1) model  (autoregressive integrated moving average) \cite{box1994} with $s=3$ \cite{anderson2, anderson5} whereas in the present work $s=6$ accurately describes the stochastic process in the absence of zonal flow component whereas by including zonal flow the ARIMA index decreases to $s=1$. Here $s$ is determined by an optimization process of the Euclidean distance between the kurtosis of the time trace and the kurtosis of the model time trace. We show that an ARIMA(1,0,1) model is sufficient to capture deterministic and stochastic components of the simulated data for the HW system with imposed energy equipartition between drift turbulence and a zonal flow component.

Theoretical models have previously successfully predicted the functional form of the PDF tail for the electrostatic fluctuations  described by drift turbulence models. Note that most of the PDFs admits a Gram-Charlier expansion, i.e. expansion in a known distribution, usually the Gaussian, indicating that small fluctuations are close to a Gaussian. In comparing with theoretical models we utilize analytical results from nonperturbative stochastic theory, the so-called instanton method \cite{justin1989, gurarie1996, falcovich1996, kim2002, anderson1, kim2008, anderson2} for computing PDFs. The analytically derived PDFs are rather insensitive to the details of the linear physics of the system \cite{kim2008} and thus display salient features of the nonlinear interactions or the details of a coupled field theory. The numerically generated time traces are analysed using the ARIMA model and subsequently fitted with the analytical models accordingly. We find in the regions with strong nonlinear characteristics an emergent universal scaling of the PDF tails of exponential form $\sim\exp\big(- const \ |\phi|\big)$ as suggested by recent theoretical work in Ref. \onlinecite{falcovich2011, kim2008, anderson4} relevant for the direct cascade dynamics. As is suggested by the ARIMA model with $s=1$ we are close to randomization due to the interaction between the drift waves and the zonal flows. However, in many cases we find Gaussian PDFs for small events. The HW model is described in section (\ref{sec:HW}) and a statistical analysis with interpretation is performed in section (\ref{sec:SA}). The remainder of the paper is devoted to the quantification of intermittency and then finally a discussion. 


\section{Hasegawa - Wakatani model}
\label{sec:HW}
The Hasegawa-Wakatani equations (HWE) provide a conceptual model of transport processes in magnetically confined (MCF) plasma
\cite{hmk1979,Scott88}. Numerical simulations based on these equations capture the key elements of MCF plasma dynamics:
drift instability due to non-adiabatic electron response, onset of drift turbulence and the self-organisation of plasma
into zonal flows. The HWE offer an alternative to quantitatively superior, but increasingly complex models,
such as gyrokinetics.

The HWE describe low frequency ($\omega << \omega_{ci}$, where $\omega_{ci}$ is the ion gyro frequency) fluctuations of
the density $n$ and the electrostatic potential $\phi$, in the presence of the constant background density gradient,
parallel electron resistivity and for a small ion-electron temperature ratio ($T_i/T_e << 1$). In the presence of
axisymmetric zonal flows with poloidal wave number $m=0$, which do not contribute to the parallel currents, the HWE are
in the quasi two dimensional form:
\begin{eqnarray}
\frac{\partial n}{\partial t} &=&-\kappa \frac{\partial \phi}{\partial y}+\alpha (\tilde{\phi}-\tilde{n})+[n,\phi]+D \nabla^2 n
\label{HW1}\\
\frac{\partial}{\partial t} \nabla^2\phi &=&\alpha (\tilde{\phi}-\tilde{n})+[\nabla^2\phi,\phi]+\mu \nabla^2 (\nabla^2 \phi), 
\label{HW2}
\end{eqnarray}
where total fluctuating fields $n$ and $\phi$ were reconstructed into turbulent parts, $\tilde{n}$, $\tilde{\phi}$, and
zonal fluctuations $\langle n \rangle$ and $\langle \phi \rangle$, that is, $n=\tilde{n}+\langle n \rangle$ and 
$\phi=\tilde{\phi}+\langle \phi \rangle$. The square brackets $\langle \dots \rangle$ indicate poloidal averages, which
in the slab model simply indicate integration along the poloidal line at a given radial location:
\begin{equation}
 \langle f \rangle= \frac{1}{L_y} \int_0^{L_y} f dy.
\label{fsa}
\end{equation}

The nonlinear advection terms are expressed as Poisson brackets
$[A,B] = \partial A/\partial x .\partial B/ \partial y - \partial A/\partial y .\partial B/\partial x$. In both equations
physical quantities have been normalized using $e\phi/T_e \to \phi$, $n/n_0 \to n$, $\omega_{ci} t \to t$
and $(x,y)/\rho_s \to x,y$. Standard notation is used for other quantities: $T_e$ is the electron temperature,
$\omega_{ci}$ is the ion gyrofrequency, and $\rho_s = \sqrt{m_i T_e}/eB$ is the hybrid Larmor radius. Dissipation terms
of the form $\nabla^2 \phi$ are added to the equations for numerical stability, where $D$ and $\mu$ are dissipation coefficients.
There are physical interpretations for these coefficients where $D$ is identified with the cross-field ambipolar diffusion and
$\mu$ is the ion perpendicular viscosity.
The $x$ and $y$ directions are identified, respectively, with the radial and poloidal directions in a tokamak, and the
magnetic field is assumed to point in the $z$ direction. We assume that $\kappa = -\partial \ln(n_0)/\partial x$ determines
the background density profile $n_0(x)$. The parameter $\alpha$ controls the strength of the resistive coupling between
$n$ and $\phi$ through the parallel current,
\begin{eqnarray}
\alpha = \frac{T_e k^2}{n_0 e^2 \eta \omega_{ci}}
\end{eqnarray}
where $\eta$ is electron resistivity. The adiabaticity parameter $\alpha$ determines the degree to which electrons can move
rapidly along the magnetic field lines and establish a perturbed Boltzmann density response. The HW system approaches
different regimes in the limits of $\alpha \to 0$ and $\alpha \to \infty$. When $\alpha \to \infty$ and $n \to \phi$ (that
is density fluctuations become enslaved to the electrostatic potential fluctuations) the HWE become identical to a one field
indirectly forced Charney-Hasegawa-Mima equation \cite{hm1977,hm1978}. In the limit $\alpha \to 0$ the equation becomes 
equivalent to the incompressible Euler equation in 2D.
Here a non-zero $\alpha$ leads to a growth rate yielding a random stirring that prevents the vorticity to decay to zero as is 
the usual case with the unforced Navier-Stokes. We note that there are both stable and unstable waves present in the system. 

Equations \eqref{HW1} and \eqref{HW2} were solved numerically on the square grid of size $L=40$ (units of $\rho_s$) with 
spatial resolution of 
$256\times256$ grid points. It is well known that the final stage of the HW evolution is dominated by zonal flows. In 
real laboratory plasmas and in 3D simulations, the poloidal damping of zonal flows enforces a certain level of energy
equipartition between turbulence and zonal flows. We enforce such equipartition in our simulations applying the following
algorithm. We modify the decomposition of fluctuations so that each fluctuating quantity $x=\tilde{x}+\gamma \langle x \rangle$
($x=n, \phi$), that is we multiply each poloidally averaged zonal flow component by factor $\gamma$.
In each step of the simulation we monitor the kinetic energy of zonal flows
\begin{equation}
 \langle E \rangle_K \equiv \frac{1}{2} \int \left( \frac{\partial \langle \phi \rangle}{\partial x} \right)^2 dV,
\label{ZKE}
\end{equation}
and its turbulent counterpart, given by
\begin{equation}
E_K \equiv \frac{1}{2} \int \left( \nabla \phi \right)^2 dV.
\label{TKE}
\end{equation}
If at any time step $\langle E \rangle_K > E_K$, we set $\gamma=E_K / \langle E \rangle_K$, otherwise we set $\gamma=1$
\cite{dewhurst2009}.

We performed three simulations varying the parallel coupling coefficient $\alpha=0.25$, $0.5$ and $1.0$, which capture
distinct dynamical regimes for the HWE. We examine poloidally averaged potential, vorticity and density flux, which is
calculated by integral \eqref{fsa} of the point wise density flux:
\begin{equation}
 \Gamma_y=\langle \Gamma \rangle= \frac{1}{L_y} \int_0^{L_y} n \frac{\partial \phi}{\partial y} dy.
\label{fsaGamma}
\end{equation}
The data is collected in ten evenly spaced radial location starting from location $x=20$ and ending at $x=200$.

\section{Statistical analysis}
\label{sec:SA}
\subsection{Spectra analysis of averaged particle flux}
\begin{figure}[ht]
\includegraphics[width=5.7cm, height=5.6cm]{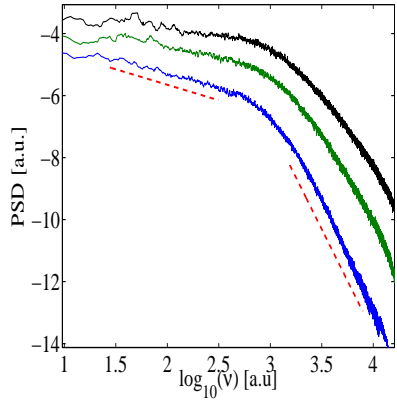}
\caption{The power spectra of averaged flux for $\alpha = 0.25$ (blue), $\alpha = 0.5$ (green) and $\alpha = 1.0$
(black).}
\label{FluxPSD_Compared}
\end{figure}

Particle flux, integrated over a flux surface, is one of the main quantities of interest for experimental
magnetically confined plasmas. Statistical features of these signals are heavily influenced by the presence of
long lived coherent structures of different scale size. The low frequency region ($\omega \sim \omega_d$, where 
$\omega_d$ is a drift frequency) of the power spectrum is of particular interest, as it corresponds to the
large flux events. Figure \ref{FluxPSD_Compared} shows power spectra from three runs with different parameter
$\alpha$. These spectra were averaged over all available ensembles measured at different radial locations.
A dashed line corresponding to the $P(\nu) \sim \nu^{-1}$ scaling has been provided to guide the eye.
These spectra clearly show different behaviour in the low frequency limit, which here we take as $\nu < 10^{3}$.
For the smallest value of $\alpha=0.25$ we see a convincing $P(\nu) \sim \nu^{-1}$ scaling region, extending
approximately a decade, for $10^{1.5}< \nu < 10^{2.5}$. The spectra then steepen considerably for large frequencies.
In the same low frequency range, the spectrum for $\alpha=1$ is nearly flat, with the best linear fit (for the
logarithmic values) giving a slope of about $0.4$.

There is no unique explanation of the origin for the $\nu^{-1}$ scaling of the power spectrum, such as observed
for $\alpha=0.25$. Two alternative dynamical systems, often discussed in the literature, are turbulence with
strong coupling to shear scales and avalanching transport, as observed in nonlinear systems near critical threshold.
In hydrodynamic turbulence with the background velocity shear
larger than the velocity shear of turbulent fluctuations, the distortion of turbulent structures by the background
shear dominates over distortions due to nonlinear interactions. In such cases, a new scaling regime in the power spectrum,
$P(\nu) \sim \nu^{-1}$ is generated \cite{Tchen1953}. Naively, one would expect that this scaling should appear
for $\alpha=0.25$, since in the limit of $\alpha << 1$ the HWE approach the dynamics of the 2D hydrodynamic turbulence.
In this limit the density perturbation is passive and simply maintain the incompressibility condition, thus the
flux fluctuations may reflect those of radial velocity. However, avalanche-like dynamics of flux events has also been
observed in numerical simulations, even if the spectra did not exhibit $\nu^{-1}$ scaling \cite{McMillan2009}.

For large values of $\alpha$ the system resembles the dynamics of the Hasegawa-Mima equation. It is known that
in this case the parallel dynamics controls the size of the largest vortices generated at small wave number (and
thus small frequencies). It is also well known that for the system of structures with the finite average size $\tau_s$
the spectrum will have an approximately Lorentzian shape with the flat low frequency region extending to frequencies
$\nu_{s} \sim (2T_s)^{-1}$. These structures do not interact strongly with others, and this leads to the change in
the power spectrum scaling.

One interpretation of the emergence of power-law scaling of the power spectrum can be found by studying the autocorrelation
function of the particle diffusion,
\begin{equation}
R^{\Gamma}(\tau) = \int_{-\infty}^{\infty} \Gamma(t) \Gamma^{*}(t+\tau)dt
\end{equation}
where $\Gamma^{*}(t)$ is the complex conjugate of the diffusion. We can now employ the Wiener-Khinchin theorem to find the power
spectrum of the form $|\Gamma(\omega)|^2$ determined by the Fourier transform
\begin{equation}
|\Gamma(\omega)|^2 = Re \int_{-\infty}^{\infty} R^{\Gamma}(\tau) e^{-i \omega \tau} d \tau.
\label{eq:int}
\end{equation}
For purely diffusive transport the autocorrelation function usually decays exponentially $R^{\Gamma} (\tau) \sim e^{-\Delta \omega \tau}$
where $\Delta \omega$, the inverse of the turbulence correlation time, can be approximated by the mixing length result
$\Delta \omega \sim k^2 D$. This gives the approximate scaling of the form:
\begin{eqnarray}
|\Gamma(\omega)|^2 \sim \frac{2 k^2 D}{(k^2 D)^2 + \omega^2}.
\end{eqnarray}
This simple estimate gives in the limit $\omega << (k^2 D)$ the result $\Gamma \sim \omega^0$ whereas in the large frequency limit
$\omega >> (k^2 D)$ one obtains $\Gamma \sim \omega^{-1}$. However the scaling might easily deviate from this estimate and also the
autocorrelation may not decay in an exponential manner, instead we may assume the form:
\begin{equation}
R^{\Gamma}(\tau) \sim e^{-\Delta_{\beta} \tau^{\beta}}.
\end{equation}
According to Ref \onlinecite{anderson2014} we find that the integral \eqref{eq:int} can be replaced by the $E_{\beta}$-function,
\begin{equation}
|\Gamma(\omega)|^2 \sim \frac{a}{(1 + b(q-1)\omega^2)^{1/(q-1)}},
\end{equation}
where $\beta$ and $q$ are related through $\beta = (3-q)/(q-1)$ and $q = (3+\beta)/(1+\beta)$ for fractional $\beta<2$.
We find an indication that there might be intermediate power law scalings that can be obtained by assuming slight deviations from
the regular diffusion picture. Moreover the precise value of $\beta$ can be determined by fitting the tails. Note that
this solutions corresponds to the Fokker-Planck equation with a fractional derivative in velocity space \cite{anderson2014}.
 
\subsection{Quantifying intermittency}
Before we proceed with quantifying intermittency in the simulated time traces from the HW system we introduce the theoretical framework used for this interpretation. We will quantify the intermittency in the simulated time series by computing the PDFs of the 
residuals or the stochastic component of the time traces and compare these with analytical predictions. Note that it is predominantly
the large scale events such as those mediated by coherent structures that contribute to the intermittency. Here, we briefly outline the 
implementation of the instanton method used in computing the tails of the distribution function. For more details, the reader is
referred to the existing literature \cite{justin1989}. A general class of solutions for the scaling of the PDF tail is presented
in Ref. \onlinecite{kim2008}. In the instanton method the PDF tail is first formally expressed in terms of a path integral by
utilizing the Gaussian statistics of the forcing, in a  similar spirit as in Refs. \onlinecite{justin1989, kim2002, anderson1, kim2008, 
anderson2}. Here and throughout this paper, the term forcing is meant to describe the inherent unpredictability of the dynamics
and will be assumed to be Gaussian for simplicity. The integral in the action ($S_{\lambda}$) in the path integral is evaluated
using the saddle-point method in the limit $\lambda \rightarrow \infty$ representing the tail values. The parameter $\lambda$ is 
proportional to some power of the quantity of interest such as the potential or flux. In mathematical terms, this corresponds to 
evaluating the integral along an optimum path described by the instanton among all possible paths or functional values. The
instanton is localized in time, existing only during the formation of coherent structure. We approximate the contribution of the 
instanton by the saddle-point solution of the dynamical variable $\phi({\bf x},t)$ of the form $\phi({\bf x},t)=F(t)\psi({\bf x})$.
We use the initial conditions for the instanton as $F(t) = 0$ at $t=-\infty$ and $F(t) \neq 0$ at $t=0$. Note that the function
$\psi({\bf x})$ here represents the spatial form of the coherent structure. Thus, the intermittent character of the transport
consisting of bursty events can be described by the creation of the coherent structures. The dynamical system with a stochastic
forcing is enforced to be satisfied by introducing a larger state space involving a conjugate variable $\phi^*$, whereby $\phi$
and $\phi^*$  constitute an uncertainty relation. Furthermore, $\phi^*$ acts as a mediator between the observables (potential
or vorticity) and instantons (physical variables) through stochastic forcing. Based on the assumption that the total PDF can be 
characterized by an exponential form and that it is symmetric around the mean value $\mu$, the expression
\begin{eqnarray}
P(\phi) & = & \frac{1}{Nb} \exp{\{ - \frac{1}{b} |\phi-\mu |^{\chi}\}},
\label{pq2}
\end{eqnarray}
is found, where the potential $\phi$ plays the role of the stochastic variable, with $P(\phi)$ determining its statistical properties. 
Here $b$ is a constant containing the physical properties of the system. Using the instanton method we find different scalings of
the PDF tails that are determined by the nonlinear dynamical equation. In a vorticity conserving system the intermittent properties
of the time series in simulations are attributed to rare events of modon like structures that have a simplified response for the 
vorticity,
\begin{eqnarray} \label{modrel}
\nabla_{\perp}^2 \phi = - k_{\perp}^2 \phi + \eta x.
\end{eqnarray}
Here we note that the modon solutions is applicable for the HW system for $\eta = 1 + (1 - k^2)U$ and the vortex speed is $U$.
In this situation it has been predicted \cite{falcovich2011, anderson4} that the system has exponential tails in the direct cascade, 
$\exp\big(- const \ |\omega|\big) \sim \exp\big(- const \ |\nabla_{\perp}^2 \phi|\big)  \sim \exp\big(-const \ |k_{\perp}^2 \phi|\big)$, 
indicating a value of $\chi=1.0$ as in Ref. \onlinecite{falcovich2011, anderson4}. 
In References \onlinecite{kim2002, anderson1, anderson2} the statistics of the momentum flux is found to be a stretched exponential 
with $\chi=3/2$. However, when the nonlinear interactions are weak, as well as in the case of an imposed zonal flow, we find Gaussian 
statistics where $\chi=2$ as is elucidated in Ref. \onlinecite{anderson3}. In the analysis we will make use of different types of 
distributions to retro-fit the PDFs of simulation results mainly using the Laplace distribution ($\chi = 1.0$) and  the Gaussian 
distribution ($\chi = 2.0$).

We focus on the time traces (averaged in the poloidal $y$-direction) at five fixed radial points located at $x=40, 80, 100, 140, 180$.
Each set of data describes the time evolution of the potential and vorticity to which we apply a standard Box-Jenkins modelling~\cite{box1994}. This mathematical procedure effectively removes deterministic autocorrelations from the system, allowing for the statistical interpretation of the residual part, which a posteriori turns out to be relevant for comparison with the analytical theory. The particular model used is a subset of a general class of models called ARIMA(p,d,q) used to model a time series. The number of lagged values are $p$ and represents the autoregressive part of the model (AR) and $q$ is the number lagged values of the error term which in turn represents the moving average (MA) part of the model. The index $d$ is the number of times the data has to be differenced to achieve a stationary time series. Note that if no differencing is needed the ARMA and ARIMA models coincide, which is the case in the present analysis. The ARIMA model is here found by minimizing the Euclidean distance ($d$) between the kurtosis of the original time series and that of the ARIMA modelled time trace. The Euclidean distance is defined here as,
\begin{eqnarray}
d = \sqrt{\sum_{x=1}^{10} (K_x^{O}-K_x^A)^2},
\end{eqnarray} \label{dist}
where $K_x^{O}$ and $K_x^A$ are the kurtosis at $x$ for the original and the ARIMA modelled time traces.
\begin{figure}[ht]
{\vspace{2mm}
\includegraphics[width=5.7cm, height=5.6cm]{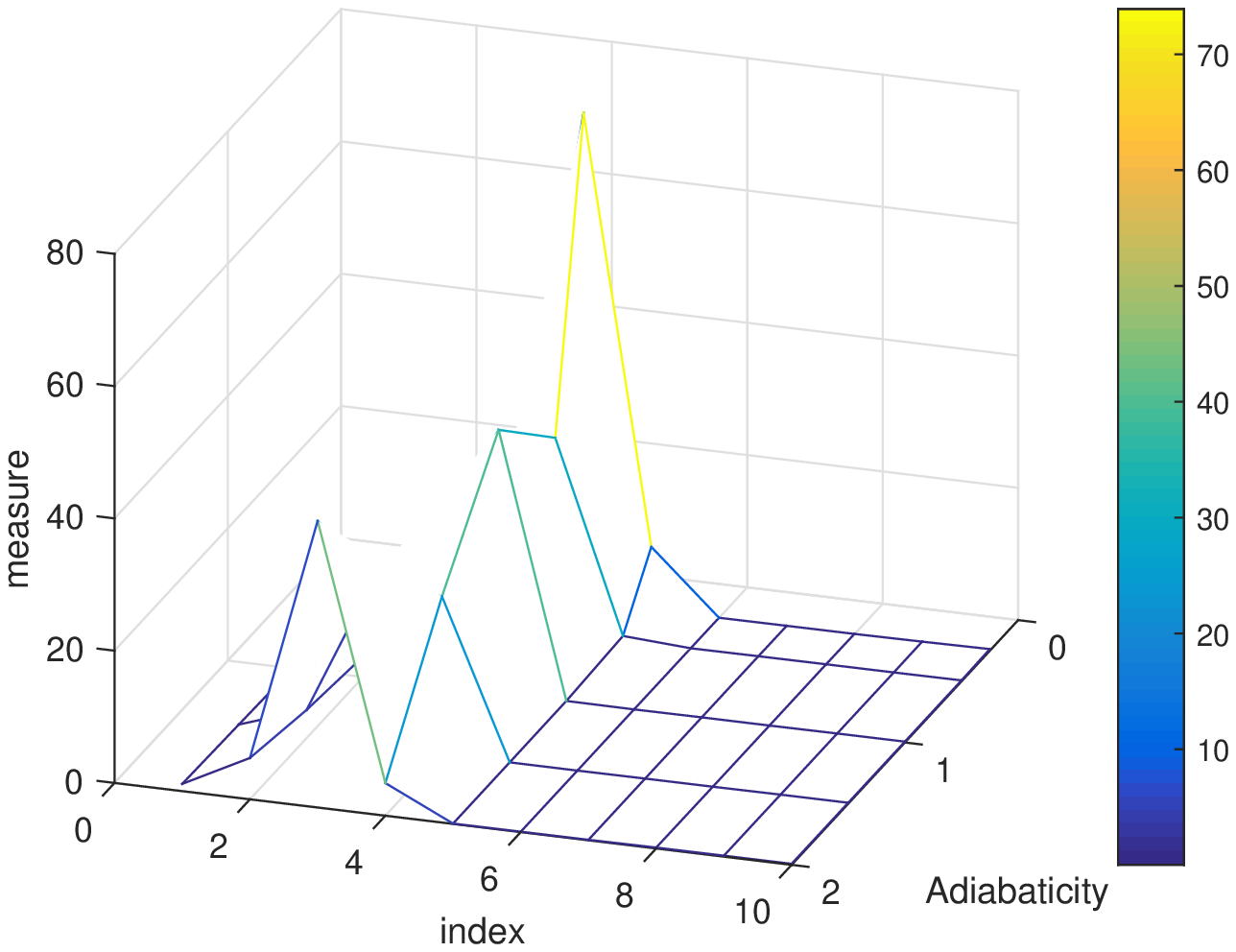}}
{\includegraphics[width=5.5cm, height=5.5cm]{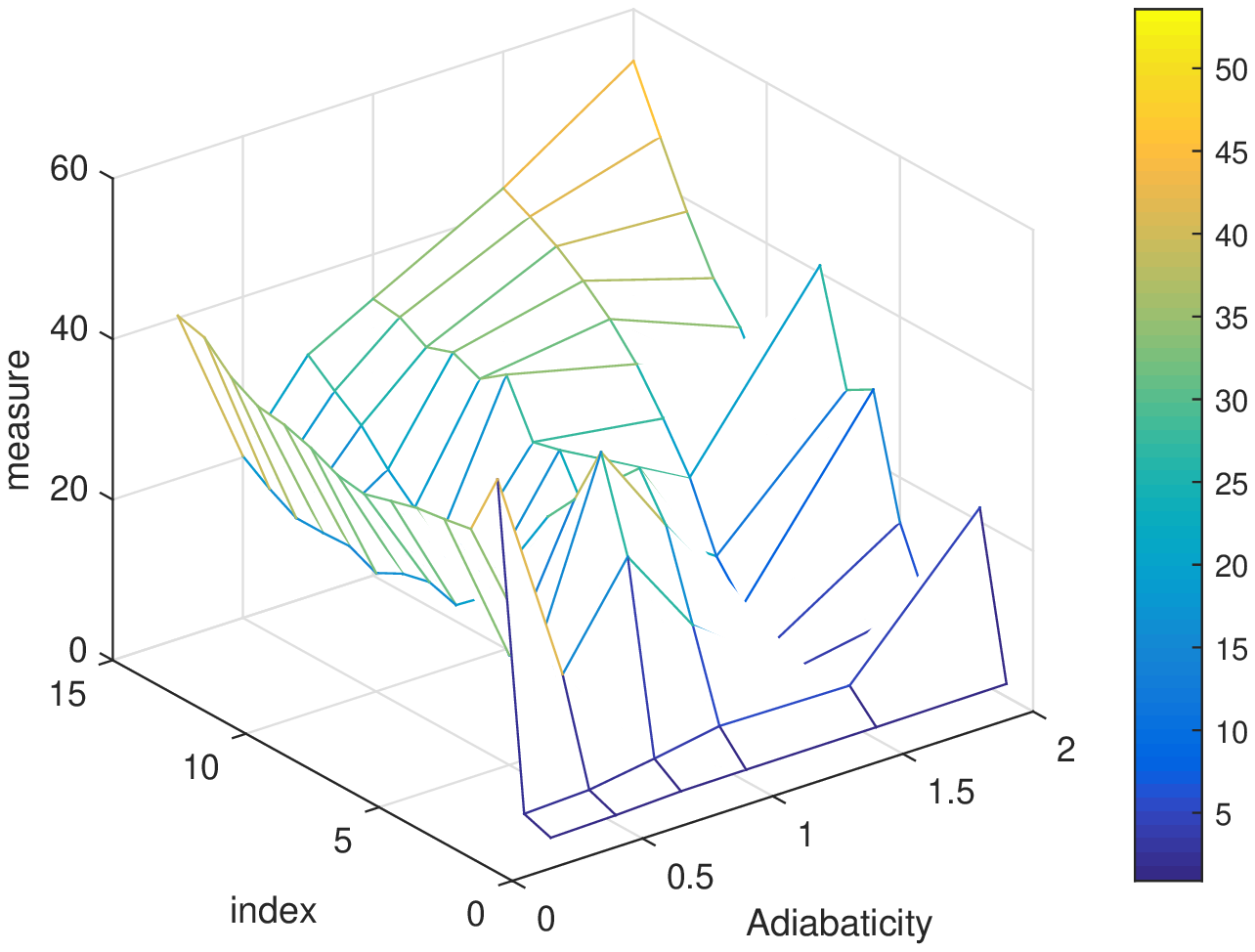}}
\caption{
The Euclidean distance (measure) as a function of the ARIMA index and adiabaticity ($\alpha$) for the case without zonal flows
(left) and with zonal flows (right).}
\label{f:measure}
\end{figure}
The ARIMA index is here determined as the lowest number yielding minimum Euclidean distance for all values of the adiabaticity
parameter. We allow for different values of $n$ in cases with and without zonal flow since the physical processes that determines
the turbulence are changed. In our set-up, it turns out that an ARIMA(n,0,1) ($n$ is the number of time lags and $1$ is the
number of noise terms) model accurately describes the stochastic process. In this case one can
express the (differenced) potential time trace with or without zonal flows in the form,
\begin{eqnarray}
\phi_{t+1}= \sum_{i=0}^{n} a_i\,\phi_{t-i}+\phi_{res}(t)
\end{eqnarray}
where the fitted coefficients $a_i$ describe the deterministic component and $\phi_{res}$ is the residual part (noise or stochastic 
component). According to Figure \ref{f:measure} we find that for the case without zonal flows $n=6$, whereas with zonal flows $n=1$. Note 
that in the case of a random walk we find that a differencing scheme with $n=1$ is sufficient to determine the dynamics. This suggests 
that the zonal flows shear larger eddies and randomizes the turbulence. In addition to be able to separate coherent structures and the random turbulent structures by the application of the ARIMA model, the index $n$ indicates the correlation length in the time series.  Thus we also find that the zonal flows significantly decreases the correlation time by randomizing turbulent structures not only in space but in time as well. Here we will mostly consider the case including zonal flows 
since neglecting zonal flows is un-physical restriction to the model. We assume that the flux is originating from the same physics that 
describes the potential $\phi$ and the vorticity $\nabla^2 \phi$, thus we keep the same ARIMA model in the analysis of the flux time 
traces. The original simulation data sets are down-sampled and consists of typically $5\times 10^5$ entries. Thus, we employ the same 
model for the electrostatic potential, vorticity and the flux for all values of the adiabaticity parameter ($\alpha$) from 0.25 to 2.00. 
We present all cases in turn with increasing $\alpha$.
\newpage
\section{Intermittency in time traces of potential and vorticity with zonal flow}

\begin{figure}[ht]
{\vspace{2mm}
\includegraphics[width=5.7cm, height=5.6cm]{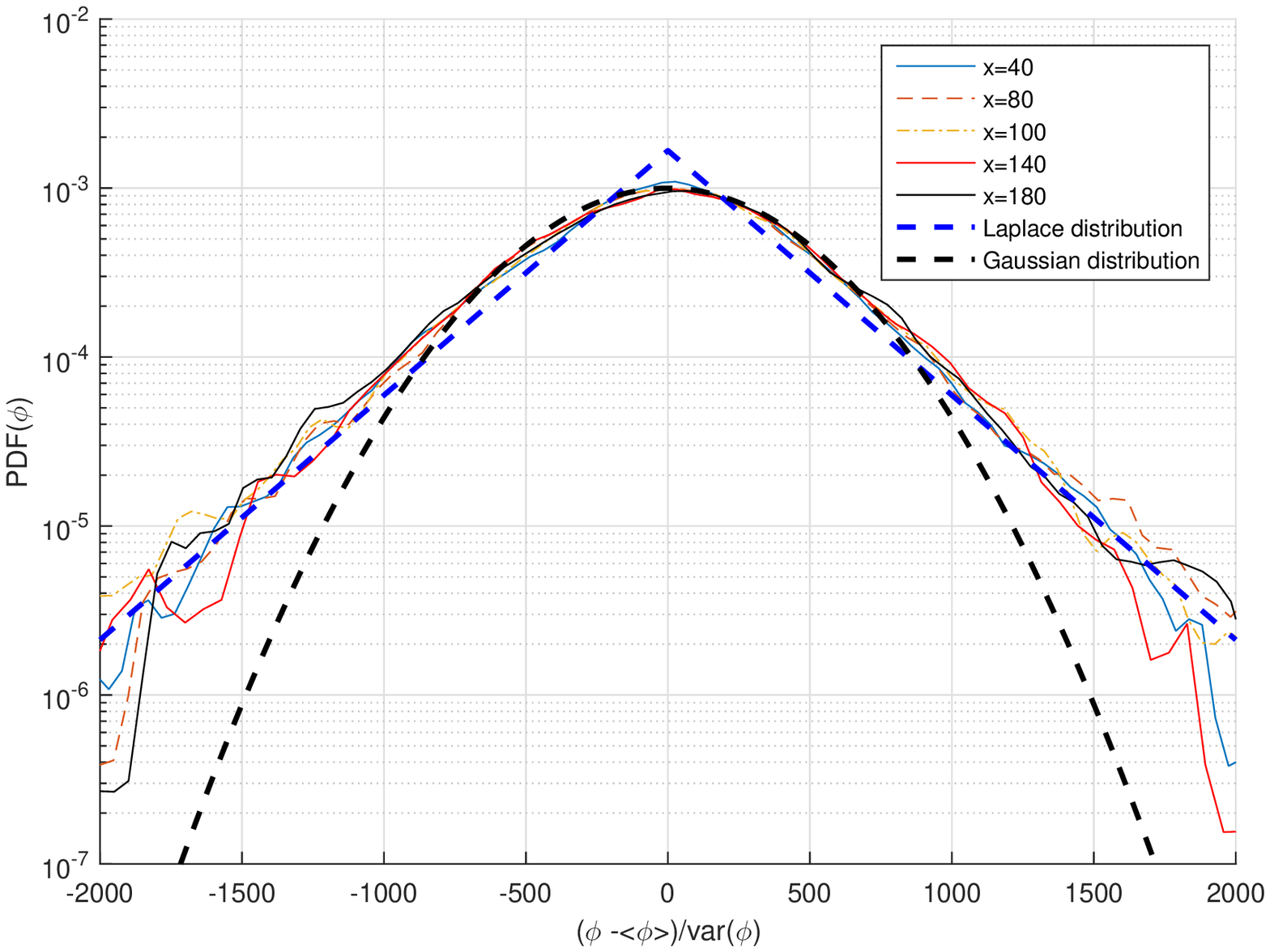}}
{\includegraphics[width=5.5cm, height=5.5cm]{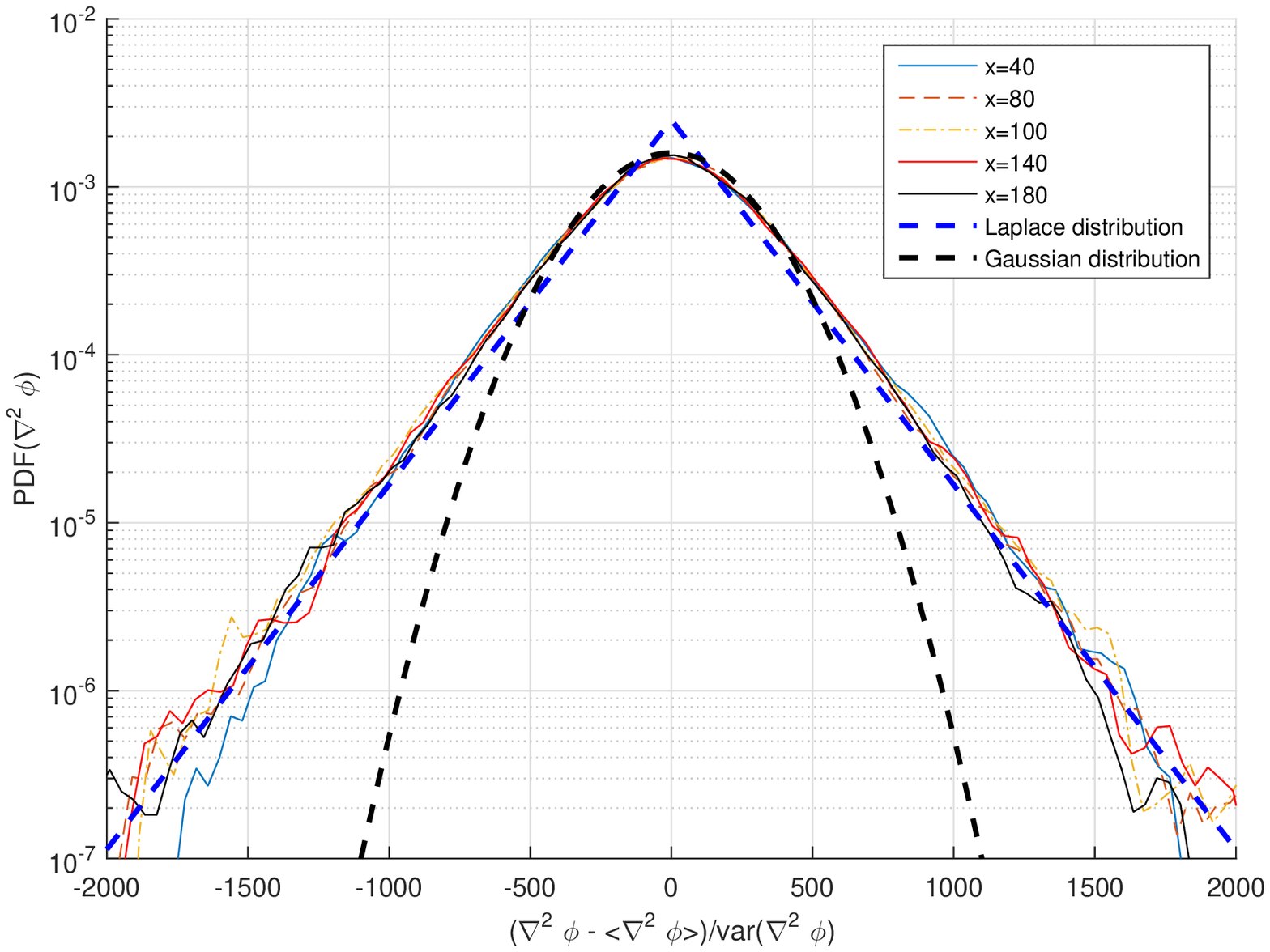}}
\caption{
The PDFs of potential and vorticity at a few radial positions $(40, 80, 100, 140, 180)$ for $\alpha = 0.25$ normalized by the variance of the individual time traces with Laplacian and Gaussian fits.
}
\label{f:A025_pdfs}
\end{figure}

\begin{figure}[ht]
{\vspace{2mm}
\includegraphics[width=5.7cm, height=5.6cm]{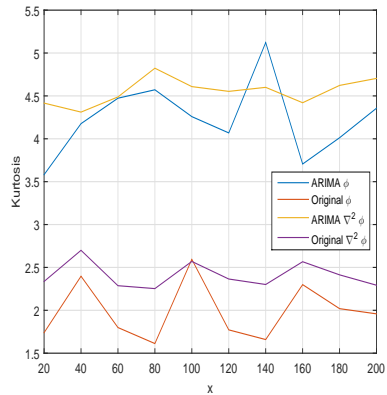}
}
\caption{
The kurtosis of the potential ($\phi$) and vorticity ($\nabla^2\phi$) time traces along the x direction for $\alpha = 0.25$.
}
\label{f:A025_kurtosis}
\end{figure}

Time traces of potential and vorticity along the x-coordinate (radial direction) are analysed using the ARIMA(1,0,1) model and the PDFs of the stochastic component obtained from this method are shown in Figure \ref{f:A025_pdfs}. Note, that for ease of identification of different radial positions, only a subset of PDFs are shown and that we have normalized the PDFs with the variance. We find a well converged PDF over several standard deviations of fluctuations. In addition, we perform retro-fitting with Laplace ($\chi = 1.00$ ) and Gaussian ($\chi = 2.0$) distributions as was done in Ref. \onlinecite{anderson5}. In Figure \ref{f:A025_pdfs}, we find tails that are significantly elevated compared to the Gaussian distribution indicating that intermittent events with large amplitude are present.

Figure \ref{f:A025_kurtosis} shows kurtosis along the x-coordinate for the original potential and vorticity signals as well as their stochastic components obtained from ARIMA method. Recalling that kurtosis for the normal distribution is equal to $3$, we observe that the original fluctuations show a sub-diffusive character, while their stochastic components show super-diffusion. The sub-diffusive character of the original fluctuations is also reflected in their PDFs, which have suppressed tails for large fluctuations. This suggests that the original time series are strongly dominated by slowly evolving, passively advected potential and vorticity structures, for which transient (across the simulation box) time is much shorter than their evolution (diffusion) time. It is only after the application of the ARIMA model that the stochastic part of the flow can be recovered and these stochastic fluctuations are super-diffusive. This suggests that turbulent eddies are strongly influenced by the zonal flow component, for example, via shearing and then viscous dissipation. This could be due to several reasons e.g. in Ref.\onlinecite{anderson5} zonal flow dynamics of the Charney-Hasegawa-Mima model where the zonal flow dynamics influences the turbulent eddies. 

For increasing values of the adiabaticity $\alpha = 1.00$ and ultimately $\alpha = 2.0$ we find similar resulting PDFs as we found for smaller $\alpha$ in Figure \ref{f:A025_pdfs} and kurtosis Figure \ref{f:A025_kurtosis} however there is a clear indication that PDFs with $1 < \chi < 2$ exists, not explicitly shown. Remembering that PDFs of this type also allows for large scale events mediating transport. We note that the center part of the PDF can be captured by a Gaussian distribution, however large elevated tails are present that cannot be fitted by a Gaussian distribution. Furthermore, we note that for $\alpha < 2.0$ a similar trend as shown in Figure \ref{f:A025_kurtosis} is found. However for $\alpha = 2.0$ it is clear that there is a deviation in the kurtosis profiles at some radial positions and only a partial match in kurtosis can be provided by the ARIMA model for potential and vorticity.

\section{Fluxes influenced by zonal flow}
We will now consider the fluxes influenced by the zonal flow component that is included in the simulations, with varying adiabaticity index. A similar ARIMA model is assumed and in this case the index is $1$ for all different values of the adiabaticity index. We have assumed that the physics responsible for the transport driven by the resistive drift waves are still the same thus we also keep the same ARIMA model.
\section*{The $\alpha = 0.25$ case}
\begin{figure}[ht]
{\vspace{2mm}
\includegraphics[width=5.7cm, height=5.6cm]{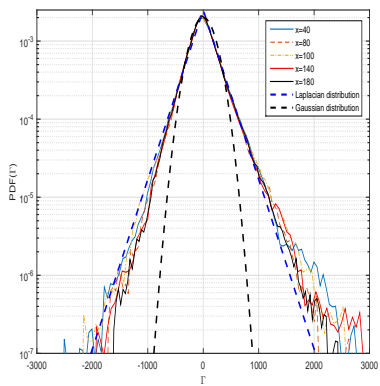}
}
\caption{
The PDFs normalized by the variance of flux at a few radial positions $(40, 80, 100, 140, 180)$ for $\alpha = 0.25$ fitted with Laplacian and Gaussian distributions.
}
\label{f:A025_pdfs_flux}
\end{figure}

\begin{figure}[ht]
{\vspace{2mm}
\includegraphics[width=5.7cm, height=5.6cm]{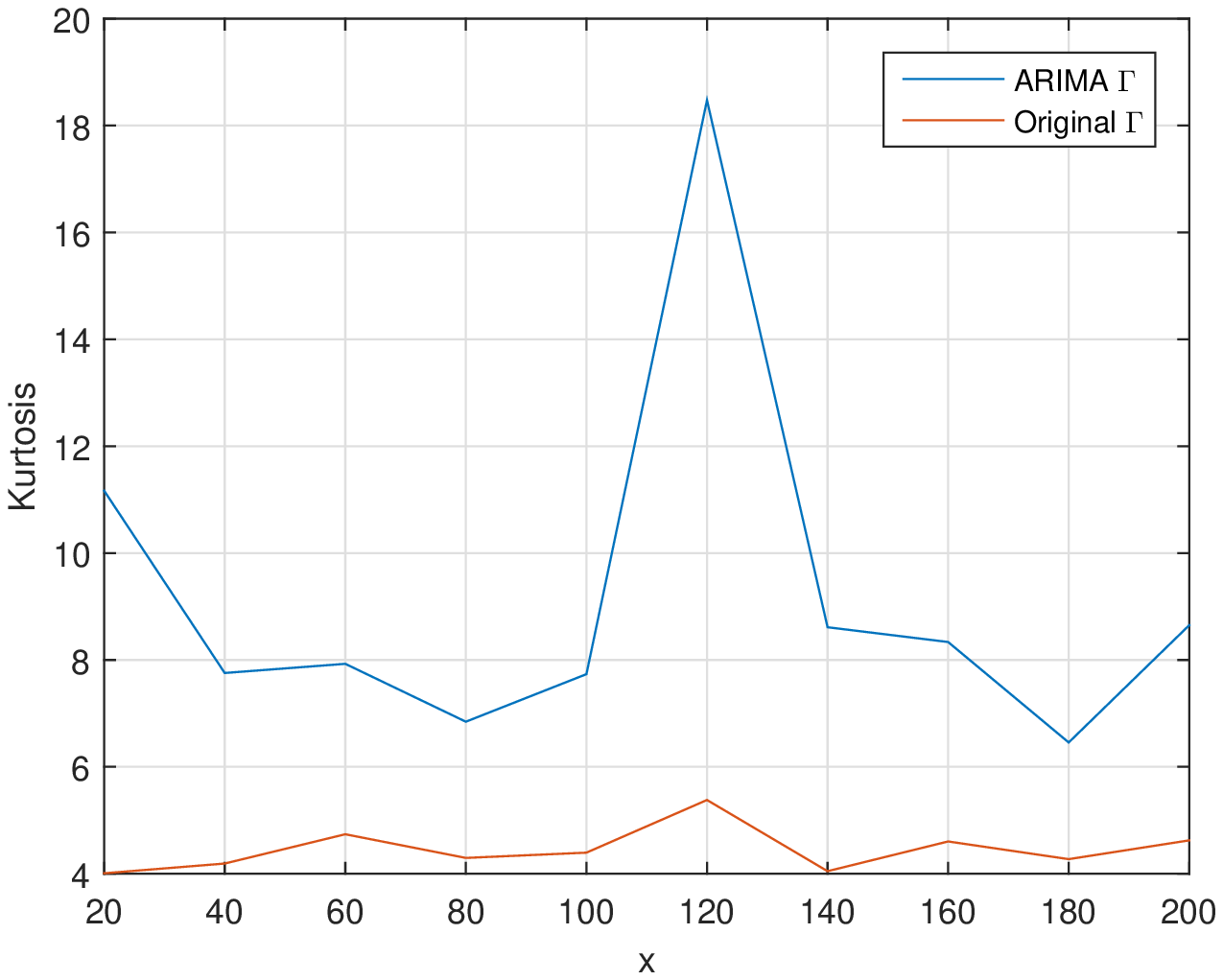}
}
\caption{
The kurtosis of the flux time traces along the x direction for $\alpha = 0.25$.
}
\label{f:A025_kurtosis_flux}
\end{figure}

\newpage
\section*{The $\alpha = 1.00$ case}

\begin{figure}[ht]
{\vspace{2mm}
\includegraphics[width=5.7cm, height=5.6cm]{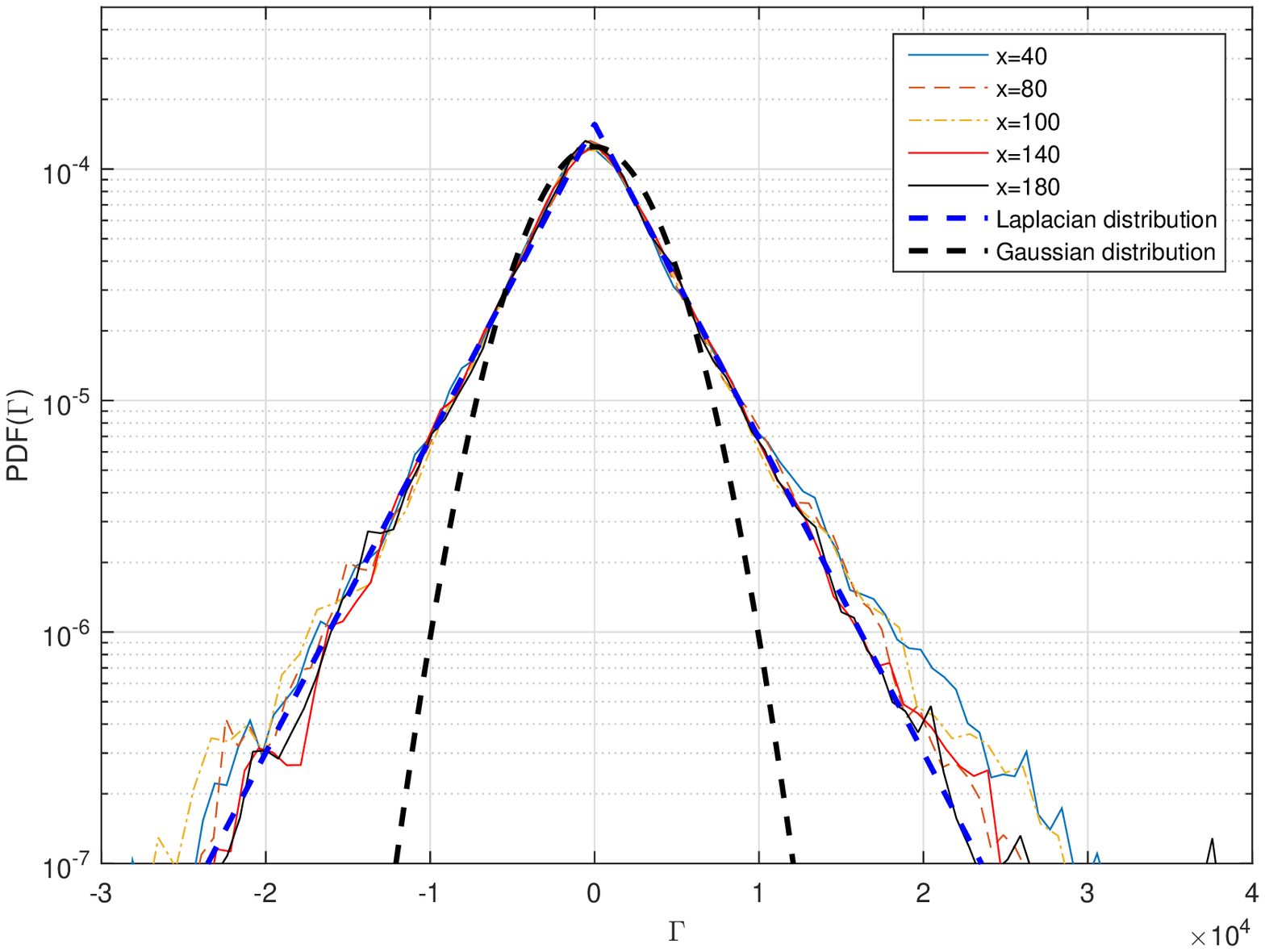}
}
\caption{
The PDFs normalized by the variance of flux at a few radial positions $(40, 80, 100, 140, 180)$ for $\alpha = 1.00$ fitted with Laplacian and Gaussian distributions.
}
\label{f:A100_pdfs_flux}
\end{figure}

\begin{figure}[ht]
{\vspace{2mm}
\includegraphics[width=5.7cm, height=5.6cm]{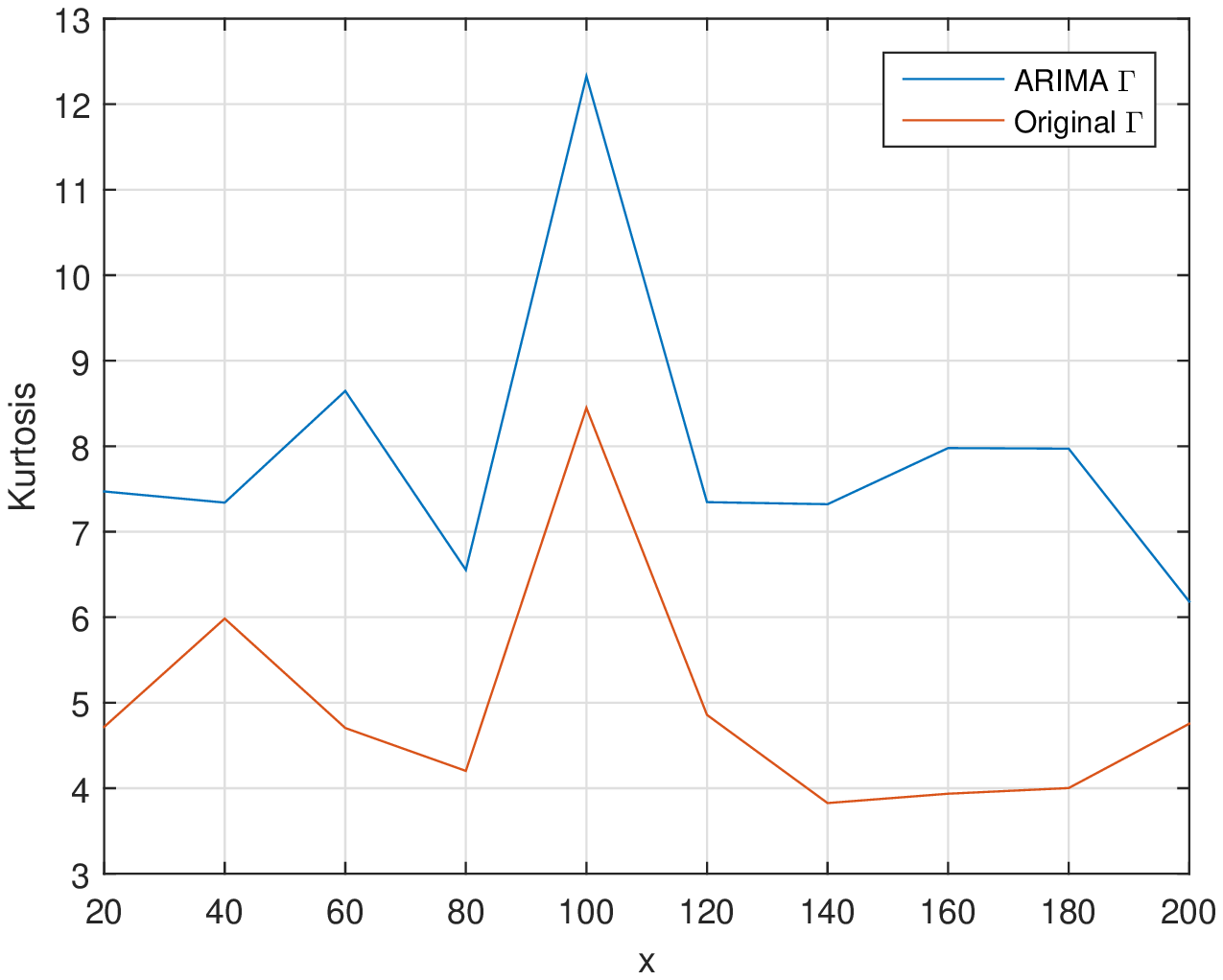}
}
\caption{
The kurtosis of the flux time traces along the x direction for $\alpha = 1.00$.
}
\label{f:A100_kurtosis_flux}
\end{figure}

\newpage
\section*{The $\alpha = 2.00$ case}

\begin{figure}[ht]
{\vspace{2mm}
\includegraphics[width=5.7cm, height=5.6cm]{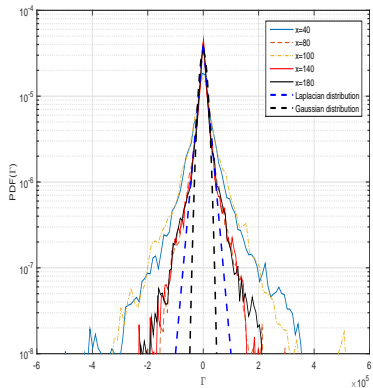}
}
\caption{
The PDFs normalized to the variance of potential and vorticity at a few radial positions $(40, 80, 100, 140, 180)$ for $\alpha = 2.00$ fitted with Laplacian and Gaussian distributions.
}
\label{f:A200_pdfs_flux}
\end{figure}

\begin{figure}[ht]
{\vspace{2mm}
\includegraphics[width=5.7cm, height=5.6cm]{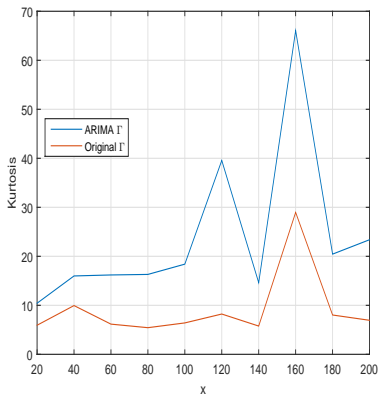}
}
\caption{
The kurtosis of the flux time traces along the x direction for $\alpha = 2.00$.
}
\label{f:A200_kurtosis_flux}
\end{figure}

In addition to the potential and vorticity time traces the fluxes are analysed using the same ARIMA model, ARIMA(1,0,1), validated by the assumption that the underlying physics responsible for the fluxes are driven by the small scale eddies that nevertheless may be impeded by the zonal flow action. There are some discrepancies in the kurtosis profiles. However the PDFs clearly display the presence of a strongly super-diffusive component, as indicated by kurtosis values elevated above $3$.  This indicates that the drift wave turbulence without zonal flows are strongly non-diffusive however the shearing action of the zonal flow induce a sub-diffusive transport state.

In general, the statistical properties change with the adiabaticity index ($\alpha$) where slightly elevated tails compared to a Gaussian is found for smaller values of adiabatic index, whereas quite large kurtosis values are found for increasing adiabatic index, as seen in Figures \ref{f:A025_kurtosis_flux}, \ref{f:A100_kurtosis_flux} and \ref{f:A200_kurtosis_flux}. Note that sub-Gaussian tails are found for $\alpha = 0.25 - 2.0$ in the original time traces. Furthermore, a quite good correspondence between model and original simulation is found for all cases studied. In all cases without zonal flows (not explicitly shown) the inherent stochastic process was close to Gaussian, however here it seems that for low $\alpha$ we find some elevated tails.  

\newpage
\section{Discussion}
We have performed a statistical analysis of the time traces for potential, vorticity and flux generated by simulations of the Hasegawa - Wakatani (HW) model with enforced equipartition of energy in large scale zonal flows and small scale drift turbulence. This simplified model describes resistive drift waves in an intermediate regime of adiabatic and hydrodynamic electrons. Nevertheless, it includes a self-consistent non-adiabatic electron response, drift wave turbulence and self-organising behaviour into zonal flows. This model allows for a qualitative understanding of the complex physics at the plasma edge in a fusion device.   

In this paper we have we have statistically analysed a number of different cases identified by different adiabatic index ($\alpha$) values that determines the strength of the resistive coupling between the density $n$ and the electrostatic potential $\phi$. The main objective for analysing this model is to characterise the statistical properties of the flux which may degrade the confinement and ultimately damage the device. Thus it is of great importance to be able to minimize large scale transport events. Previously we have seen the strong exponential character ($\chi = 1.0$) in coherent transport events at the edge in the form of meso-scale blob structures \cite{anderson4}. In order to achieve this we have performed statistical analysis of the produced time traces, averaged in the poloidal (y) - direction, at ten radial positions. The analysis is done using the ARIMA model that have been shown to efficiently remove autocorrelations, that may mask the statistical properties, from the time traces and can extract the noise component which can be tested against analytical models.

We expect that the modelled PDFs can be fitted by exponential distributions with some power $\chi$ (Eq. (\ref{pq2})) with good accuracy in most cases, however, the core of the distribution usually has a Gaussian component. We have observed that similar values of kurtosis of the potential and vorticity are recovered by the ARIMA model however the original time traces seem to have sub-Gaussian distributions, indicative of sub-diffusive behaviour where kurtosis is below three. It has previously been predicted that a diffusive character of the transport is to be expected as a result of the influence of zonal flow driven by micro-scale turbulence \cite{anderson3}. We note that there seems to be a relation between the statistics of the potential and the vorticity indication that the modon assumption in Eq. (\ref{modrel}) may be applicable, which is interesting from a modelling point of view. The physics changes significantly as the adiabaticity index ($\alpha$) is increasing. For small $\alpha$ the ARIMA modelling gives a good representation of the time traces however as $\alpha$ is increased there are some deviations in the kurtosis profiles. There are several possible reasons for this, although there is a certain randomness in the simulation results. Underneath this randomness a modon assumption is possible. The effect of the zonal flow component on the HW system is significant, it seems that without zonal flows a strong randomization of the eddies is present c.f. previous work in Ref. \onlinecite{anderson5} for more details. The overall sub-diffusive transport properties are induced by the strong interaction of the zonal flow through shearing action and time decorrelation. Furthermore, note that distribution functions can be constructed from non-linear invariants of the starting equations (\ref{HW1} and \ref{HW2}) namely linear combinations of $\frac{1}{2} \langle |\nabla_{\perp} \phi|^2 \rangle$, $\frac{1}{2} \langle (\nabla^2_{\perp} \phi)^2 \rangle$, $\frac{1}{2} \langle n^2 \rangle$ and $\frac{1}{2} \langle n \nabla^2_{\perp} \phi \rangle$ suggesting an almost Gaussian distribution in vorticity. The angle brackets denote spatial averaging over the domain. One of the complications in analysing the state of the plasma determined by the HW system is that there are both stable and unstable waves present in the plasma. This could also be one of the explanations of the difference in results here and in the previous work presented in Ref. \onlinecite{anderson5}. However, although this is a very interesting topic we leave it for future work since it is out of the scope of the present paper. 

\section{Acknowledgement}
One of the authors (J.A.) would like to thank Kyoto University for its hospitality where a large part of this work was carried out and the authors are grateful for valuable discussions with Dr. Gert J. J. Botha.

\newpage

\end{document}